\documentclass{PoS}

\title{Hydrodynamics of the CFL superfluid }

\ShortTitle{Hydrodynamics of the CFL superfluid }

\author{\speaker{Cristina Manuel}%
         \thanks{This work has been supported by the Spanish grants
AYA 2005-08013-C03-02 and FPA2007-60275}\\
        Instituto de Ciencias del Espacio (IEEC-CSIC)\\
 Campus Universitat Aut\`onoma de Barcelona, Facultat de Ci\`encies, Torre C5, E-08193 Bellaterra (Barcelona), Spain\\
        E-mail: \email{cmanuel@ieec.uab.es}}

\abstract{
At asymptotic high density and low
temperature quarks form Cooper pairs in a color-flavor locked (CFL) configuration.
The diquark condensates break spontaneously the baryon symmetry, and this fact makes the CFL phase also superfluid.
At low temperatures the transport properties are dominated by the contribution of the superfluid phonon,
the Goldstone boson associated to baryon symmetry breaking, in full analogy to what happens in
superfluid He$^4$.  We discuss how to derive transport properties in the ultracold regime making use of
an analogue model of gravity. We also review how this model can be used to study the
scattering of phonons with quantized vortices in a rotating system. Finally, we consider the implications
of these results in studying the rotational properties of compact stars made of CFL quark matter
}

\FullConference{8th Conference Quark Confinement and the Hadron Spectrum \\
		 September 1-6 2008\\
		 Mainz, Germany}

\begin{document}

\section{Introduction}

QCD in the asymptotic high baryonic density regime is in the  
color-flavor locked (CFL) phase \cite{Alford:1998mk}. This phase is characterized by the existence of diquark condensates
that lead to a symmetry breaking pattern that locks the gauge and the flavor symmetries,
and also spontaneously break the $U(1)_B$ symmetry associated to 
baryon number.
In this high density regime, and because of asymptotic freedom, one can compute all properties of CFL quark matter \cite{Alford:2007xm}. 
This task is not only a challenging and interesting exercise,  but it is also
needed  if one wants to look for signatures of quark matter in astrophysical scenarios.

In this talk we will focus our attention to the superfluid hydrodynamics of the CFL phase
at low temperatures. Superfluidity was first discovered in He$^4$, after cooling it down below
2.17 K, when it experiences a Bose-Einstein condensation.
 Landau  realized that the superfluidity  of He$^4$ was linked to 
the existence of a  collective  mode with a linear dispersion relation.  At very low temperatures
and velocities helium  flows without being able to create these elementary excitations or others,
and thus without dissipating energy.
We now understand that this collective mode, the so called superfluid phonon,
corresponds to the Goldstone mode associated to the spontaneous
breaking of the law of particle number conservation.

It is precisely the  spontaneous breaking of the $U(1)_B$ symmetry, with the appearance of the phonon
excitation, which makes the CFL a superfluid. For this reason  we expect that the hydrodynamical behavior
of CFL quark matter is characterized by all the peculiarities of superfluidity. 
The hydrodynamics of the CFL phase should be described by the relativistic version of 
Landau 's two-fluid model. There should be a superfluid component, describing the coherent motion of
the condensate, and a normal fluid component, where dissipative processes are allowed.
At very low temperatures the phonons give the main contribution to this normal component.
Further, when the system is rotation, quantized vortices should appear.

I will now explain some recent developments in this field, based on work
done in collaboration with Massimo Mannarelli,
 Felipe Llanes-Estrada, Antonio Dobado, and Basil Sa'd.

\section{Hydrodynamics in the CFL phase at zero temperature}

In the CFL phase the superfluid phonon is the Goldstone boson associated to the breaking of the $U(1)_B$ symmetry and it
can be introduced  as a phase of the diquark condensate.
The effective field theory  for this Goldstone boson  can be constructed from the equation of state  of normal quark matter~\cite{Son:2002zn}, or equivalently,  by integrating out all the
heavy modes from the QCD Lagrangian.

 At very high density, or quark chemical potential $\mu$, when  the coupling constant is small $g (\mu) \ll 1$, one finds \cite{Son:2002zn}
\begin{equation}
\label{L-BGB-0}
{\cal L}_{\rm eff}  = \frac{3}{4 \pi^2}
\left[ (\partial_0 \varphi - \mu)^2 - (\partial_i \varphi)^2
\right]^2 \, .
 \end{equation}

There is an interesting  interpretation of the equations of motion
associated to the superfluid phonon. Since the Lagrangian in Eq.~(\ref{L-BGB-0}) does not explicitly depend on the field $\varphi$, but only on its derivatives, the corresponding classical equation of motion  takes the form of  a conservation law. That equation together with the conservation law of the energy-momentum tensor can be viewed
as  hydrodynamical laws \cite{Son:2002zn}
\begin{equation}
\label{S-hy-1}
 \partial_\nu (n_0 v^\nu) = 0 \ , \qquad \partial_\rho T^{\rho \sigma}_0 = 0\ .
\end{equation}
Here $ n_0 =\frac{dP}{d \mu} \Big |_{\mu =\bar \mu} = \frac{3}{\pi^2} \bar \mu^3 $ is interpreted
as the baryon density, where  $\bar \mu = (D_\rho  \bar \varphi D^\rho  \bar \varphi)^{1/2}$ and
\begin{equation} \label{svelocity}
 v_\rho = - \frac{D_\rho  \bar\varphi}{\bar \mu} \, ,
\end{equation}
is the superfluid velocity  with $\bar \varphi$ the solution of the classical equation of motion. 
The energy-momentum tensor  can  be written in terms of the velocity
defined in Eq.~(\ref{svelocity}) and Noether's energy-density $\rho_0$,
 \begin{equation}
\label{S-hy-2}
T^{\rho \sigma}_0 = (n_0 \bar \mu)  v^\rho  v^\sigma - g^{\rho \sigma} P_0 = (\rho_0 + P_0)  v^\rho  v^\sigma - \eta^{\rho \sigma} P_0 \ ,
 \end{equation}
where we have written  $\rho_0 + P_0 = n_0 \bar \mu$, with $P_0$  the quark pressure evaluated at
$\bar \mu$.

\section{Heating up the CFL superfluid: the superfluid phonon contribution}

In a superfluid, the phase of the condensate allows one to obtain the superfluid velocity, but it also correponds
to the Goldstone boson field.
 It should  be possible to decompose this phase
in  two fields, the first describing the hydrodynamical variable, the second describing the quantum fluctuations associated
to the phonons, thus we write
$
\varphi (x) = \bar \varphi(x) + \phi(x)$ \cite{Manuel:2007pz,Mannarelli:2008jq}.
This splitting  implies a separation of scales -  the background field $\bar \varphi(x)$
is associated to the long-distance and long-time scales, while  the fluctuation $\phi(x)$ is
associated to rapid and short lenght scale variations.

From the low energy effective action of the system
we deduce the effective action for  the phonon field  expanding around the stationary point corresponding to the classical solution $\bar \varphi$.
 The action of  the linearized fluctuation - here the superfluid phonon - 
can be written as the action of a boson moving in a non-trivial gravity background
\begin{equation}
\label{phonon-action-2}
S[\phi] = \frac 12 \int d^4 x \sqrt{- {\cal G} } \, {\cal G}^{\mu \nu} \partial_\mu \phi\, \partial_\nu \phi \,,
\qquad
{\cal G}^{\mu \nu} =   
\eta^{\mu\nu} + \left(\frac {1}{c_s^2} - 1 \right) \bar v^\mu  \bar v^\nu \ ,
\end{equation}
where $c_s= 1/\sqrt{3}$ is the speed of sound.

Phonons can thus be viewed as quasiparticles which propagate following the geodesics defined by the
acoustic metric $ 0= {\cal G}_{\mu \nu} d x^\mu dx^\nu$. One can  construct the Liouville operator
so as to define the Boltzmann equation obeyed by the superfluid phonon distribution function $f(x,p)$ 
\cite{Mannarelli:2008jq}
\begin{equation}
\label{Blotzmman}
 p^\alpha \frac{\partial f}{\partial x^\alpha} - \Gamma^\alpha_{\beta\gamma} p^\beta p^\gamma  \frac{\partial f}{\partial
p^\alpha}   = C[f] \ .
\end{equation}
Here $\Gamma^\alpha_{\beta\gamma}$ are the Christoffel symbols associated to the metric ${\cal G}_{\mu \nu}$,
and $C$ is the collision term.

The idea of using gravity analogue models to describe hydrodynamical fluctuations was first used by
Unruh \cite{Unruh:1980cg}, and later on also applied to describe the dynamics of the phonons
 of a non-relativistic superfluid \cite{Volovik:2000ua}.
  In the CFL superfluid we have implemented these ideas starting from Eq.~(\ref{L-BGB-0})
 \cite{Mannarelli:2008jq}. It  allows us to express how the phonons propagate in the background of 
the superfluid, and express all the thermodynamical quantities in  a fully relativistic covariant way.
For example, in local equilibrium, we find that the contribution of the phonons to the pressure is
\begin{equation}
\tilde P_{\rm ph}  = \sqrt{- {\cal G}} \frac{T^4  \pi^2}{90 } \frac{1}{({\cal G}_{\mu \nu} u^\mu u^\nu)^{2}}
\end{equation}
which is expressed in terms of both the superfluid velocity $\bar v_\mu$ and normal fluid velocity $u_\mu$.
For low velocities $u^\mu \rightarrow  (1,{\bf u}_{NR})$ and $\bar v^\mu \rightarrow  (1,{\bf v}_{NR})$,
and thus ${\cal G}_{\mu \nu} u^\mu u^\nu \rightarrow c_s^2 - ({\bf u}_{NR}-{\bf v}_{NR})^2 $.
Then we recover the well-known expression for the pressure due to the
phonons in a non-relativistic superfluid.

The computation of transport coefficients can be  implemented  by considering the Boltzmann equation (\ref{Blotzmman}) with a non-vanishing  collision term. 
The collision term can be evaluated considering various   scattering processes  among superfluid phonons whose vertices can be read  from the effective Lagrangian in Eq.~(\ref{L-BGB-0}). The leading processes are  binary collisions, collinear splitting and joining processes \cite{Manuel:2004iv}. 

The phonon contribution to the shear viscosity, and the bulk viscosity associated to the normal fluid component have been computed following this procedure in Refs.~\cite{Manuel:2004iv,Manuel:2007pz,Manuel:2005hu}, finding
\begin{equation}
\label{final-all}
\eta= 10^{-4} \, \frac{\mu^8}{T^5}  \ , \qquad \zeta_2 = 0.0011 \, \frac{m_s^4}{T} \ ,
\end{equation}
where $m_s$ is the strange quark mass. These computations were done  
in the superfluid rest frame, where one takes $v^\mu =(1,{\bf 0})$, and
assuming homogeneity in the superfluid flow. This corresponds to consider in  Eq.~(\ref{Blotzmman})
vanishing Christoffel symbols, $\Gamma^\rho_{\mu \nu} =0$. The transport equation that we have presented will allow  us to compute other transport coefficients.

Let me finally remark that here we are  considering a low temperature regime, lower than all the energy
gaps or masses of other particles which are present in the CFL phase. Because CFL kaons are also light 
particles, with energy gaps in the order of the MeV, it is also possible that they might contribute to
transport phenomena also at relatively low temperatures   \cite{Alford:2007rw}.

\section{CFL superfluid in rotation and mutual friction}

The mean free path associated to the superfluid phonons  scales as $\ell \sim \frac{\mu^4}{T^5}$
 ~\cite{Manuel:2004iv}. 
For low temperatures $\ell$ exceeds easily the typical length of the macroscopic system one is considering.
For example, when $T \ll 0.01$ MeV, $\ell$ is bigger than  
the radius of a compact star, which is assumed to be of order $10$ Km \cite{Manuel:2004iv}. In that case
one should treat the phonon system as an ideal or non self-interacting
bosonic gas, rather  than as a normal fluid.

One may be tempted to say that there is no dissipation at all in this cold regime of the CFL phase.
However, in a rotating superfluid there are vortices, and they also allow for a new source of dissipation.
Vortices interact with the superfluid component through the Magnus force. But the quasiparticles conforming
the normal fluid also collide with these vortices, resulting in a sort of drag force acting on these string-like
objects. This effect is known as mutual friction, and it has been carefully studied in He$^4$, both theoretically
and experimentally.

The rotating CFL superfluid  is threaded with vortex lines whose density per unit area is $N_v = \frac{2 \Omega}{\kappa}$, where $\Omega$ is the rotation
 frequency of the body and where the quantized circulation 
is $\kappa=  \frac{2\pi}{\mu}$~\cite{Iida:2002ev}. 

Phonon-vortex scattering in a non-relativistic superfuid has been studied starting from the corresponding
hydrodynamical equations \cite{Fetter}. In the CFL superfluid it can also be studied with our gravity analogue
model. One simply has to use a vortex configuration for the superfluid velocity which enters in the
acoustic metric ${\cal G}_{\mu \nu}$ and study the corresponding gravitational scattering \cite{Mannarelli:2008je}. 


\section{R-modes and mutual friction in a CFL quark star}

Dissipative process are essential to understand the rotational properties of compact stars.
R-modes are non-radial oscillations of the star with the Coriolis force acting as the restoring force. They provide a severe limitation on the star's rotation frequency through coupling to  gravitational radiation (GR) emission~\cite{Andersson:2000mf}. When dissipative phenomena damp these r-modes the star can rotate without losing  angular momentum to GR. If dissipative phenomena are not strong enough,  these  oscillations  
will grow exponentially and the star will keep slowing down until some dissipation mechanism   can  damp the r-modes.

In Ref.~\cite{Mannarelli:2008je} we have studied the maximal critical frequency than an hypothetical
compact star made up by CFL matter might sustain in the  temperature regime where the only source of
dissipation is mutual friction, that is, for $T \ll 0.01$ MeV.
We first obtain the time scale associated to mutual friction
$\frac{1}{\tau_{MF}} \simeq 2 \times 18.1 \left(\frac{T}{\mu}\right)^5 \Omega  $,
and after comparing this with the time scales of growth for the r-modes, we find that only in the case
when the frequency of the star is of the order or less than 1 Hz, this mechanism is effective to stop the grotwh of
the r-modes. This result rules out the possibility that cold pulsars rotating at higher frequencies,
which are about of $75\%$ of the the observed ones,
are entirely made up by CFL quark matter.

\end{document}